\documentclass[fleqn,usenatbib]{mnras}

\usepackage{newtxtext,newtxmath}

\usepackage[T1]{fontenc}

\DeclareRobustCommand{\VAN}[3]{#2}
\let\VANthebibliography\thebibliography
\def\thebibliography{\DeclareRobustCommand{\VAN}[3]{##3}\VANthebibliography}

\usepackage{graphicx}
\usepackage{multirow}
\usepackage{graphicx}	
\usepackage{amsmath}	
\usepackage{placeins}
\usepackage{xfrac}

\usepackage{hyperref}
\hypersetup{colorlinks=true, linkcolor=Navy, citecolor=Navy, urlcolor=Navy}

\usepackage{color}
\definecolor{Navy}		{RGB}{  0,   0, 128}
\definecolor{MidnightBlue}	{RGB}{ 25,  25, 112}
\definecolor{yellow}   	{RGB}{255, 215,   0}
\definecolor{darkorange}{RGB}{255, 140,   0}
\definecolor{dodgerblue}{RGB}{ 30, 144, 255}
\definecolor{black}     {RGB}{  0,   0,   0}
\definecolor{dimgray}   {RGB}{105, 105, 105}
\definecolor{gray}   {RGB}{128, 128, 128}

\newcommand{\VAL}[1]{#1}
\newcommand{\fits}{\texttt{fits} }

\DeclareRobustCommand{\ion}[2]{%
  \relax
  \ifmmode
    \ifx\testbx\f@series
      {\mathbf{#1\,\mathsc{#2}}}
    \else
      {\mathrm{#1\,\mathsc{#2}}}
    \fi
  \else
    \textup{#1\,{\mdseries\textsc{#2}}}%
  \fi
 }

\title[ESPRESSO fiber-injection losses]{Accurate estimate of the ESPRESSO fiber-injection losses\\ inferred from integrated field-stabilization images}

\author[T. M. Schmidt]{
Tobias M. Schmidt,$^{1}$\thanks{E-mail: tobias.schmidt@unige.ch}\\
$^{1}$Observatoire Astronomique de l'Universit\'e de Gen\`eve, Chemin Pegasi 51, Sauverny, CH-1290, Switzerland
}
\date{Accepted XXX. Received YYY; in original form ZZZ}

\pubyear{2024}

\begin{document}
\label{firstpage}
\pagerange{\pageref{firstpage}--\pageref{lastpage}}
\maketitle

\begin{abstract}
Ground-based astronomy is unavoidably subject to the adverse effect of atmospheric turbulence, a.k.a. the seeing, which blurs the images and limits the achievable spatial resolution.
For spectroscopic observations, it leads to slit or fiber-injection losses, since not all photons distributed over the extended seeing disk can be captured.
These losses might have a very substantial impact on the overall efficiency of a spectrograph and are naturally highly variable.
Assessing the fiber-injection losses requires accurate information about the image quality (IQ) delivered by the telescope to the instrument over the course of the observations, which, however, is often not directly available.
ESPRESSO provides acquisition and field-stabilization images attached to the science data and thus offers the opportunity for a post-processing analysis.
Here, we present a novel method to infer the IQ profile and fiber-injection losses from the integrated field-stabilization images, utilizing the spill-over light that does not get injected into the fiber.
We validate these measurements against the IQ observed in the acquisition images and determine that our method delivers unbiased estimates with a scatter of \VAL{$0.11"$} for the FWHM of the profile and \VAL{$15\,\%$} in terms of fiber-injection losses.
This compares favorably to the estimates derived from either the differential image motion monitor (DIMM) or the telescope guide probe sensors and therefore represents a valuable tool to characterize the instrument efficiency and to correct raw spectra for fiber-injection losses.
\end{abstract}

\begin{keywords}
Atmospheric effects -- Instrumentation: spectrographs -- Techniques: spectroscopic -- Methods: data analysis -- Methods: observational
\end{keywords}


\section{Introduction}
\label{Sec:Introduction}

One of the main complications for ground-based astronomy is the blurring of the images introduced by atmospheric turbulence, commonly described as the \textit{seeing}. For imaging, this limits the achievable sharpness of the observations. For spectroscopy, seeing typically
does not hamper the spectral resolution but has a strong impact on the number of photons collected and therefore the efficiency.
In general, the achievable spectral resolution of a spectrograph is directly and inversely related to the size of the input aperture, i.e. the width of the slit or the optical fiber. Therefore, a small fiber is advantageous to reach high spectral resolution. On the other hand, the aperture of the spectrograph needs to be large enough to capture the majority of the light distributed over the seeing-limited point-spread function (PSF).
In particular for high-resolution echelle spectrographs, \'etendue is a premium. It relates various key quantities, e.g. telescope collecting area, aperture on the sky, spectral resolution, and wavelength range covered.
For volume, weight, and cost reasons, but also limitations on the manufacturable size of echelle gratings, the total \'etendue cannot be made arbitrarily large and compromises have to be made. Thus, for spectrographs on large telescopes that are supposed reach high resolving powers, e.g. $R = \frac{\lambda}{\Delta{}\lambda} \approx 100\,000$, often only a small sky aperture can be accommodated in the design. For instance, the Echelle SPectrograph for Rocky Exoplanets and Stable Spectroscopic Observations (ESPRESSO, \citealt{Pepe2021}), uses optical fibers with a diameter of only $1"$, comparable to the typical seeing.
Here, it immediately becomes clear that for geometrical reasons a notable part of the light distributed over the extended PSF cannot be accepted by the optical fiber, but also that the exact amount strongly depends on the seeing.
Under very good atmospheric conditions, the photons originating from a point-like science target are well confined in the focal plane of the telescope and a large fraction can be injected into the fiber while poor seeing leads to much higher fiber-injection losses. Here, one has to stress that for fiber-fed spectrographs these injection losses grow quadratically with the size of the PSF, while for long-slit spectrographs this dependence is just linear%
\footnote{Only the extension of the PSF perpendicular to the slit leads to direct losses. Photons spread-out along the slit are still captured. During data reduction, poor seeing mandates a larger extraction aperture in cross-dispersion direction and therefore implies larger noise contributions from read-out, dark, and sky-background, but these might not dominate the overall S/N.}.
Seeing therefore has a massive impact on the efficiency of fiber-fed spectrographs and in consequence on the achievable signal-to-noise ratio (S/N) in a given observing time.

Unfortunately, atmospheric seeing is highly dynamic and a particularly complicated aspect of ground-based astronomy \citep[see e.g.][]{Tokovinin2023}.
Most observatories employ some sort of seeing monitoring, but the values reported by e.g. a Differential Image Motion Monitor (DIMM) often do, for various reasons, not exactly represent what one measures with a science instrument itself.
Also, many related studies (see e.g. \citealt{Griffiths2024} for the most-recent one conducted at the Paranal Observatory) focus on the characterization of purely atmospheric properties and try to e.g. map the turbulence as function of altitude (typically described as a $C_n^2(h)$ profile). Here, the goal is to understand atmospheric turbulence and be able to better correct the image distortions using adaptive optics.
However, spectrographs like ESPRESSO, covering the visible and UV part of the spectrum, can currently not benefit from adaptive-optic corrections and will also in the foreseeable future work in seeing-limited mode.
For these, the fiber-injection efficiency depends on the size and shape of the PSF that is actually delivered at the focal plane of the telescope.
This is--also conceptually--a quite different property than what is measured by atmosperic tubulence characterizations.
Therefore, we adopt in the following the distinction between \textit{seeing} and \textit{image quality} (IQ), as outlined in detail by \cite{Martinez2010}. Here, \textit{seeing} describes an intrinsic property of the atmosphere that can e.g. be characterized with dedicated instruments (see Section~\ref{Sec:IQ_DIMM}). The IQ, on the other hand, relates to the PSF that is actually delivered in the focal plane of the telescope and identical with the flux profile of long-exposed point sources.
This should be dominated by blurring due to atmospheric turbulence, but other effects, like aberrations from the telescope or guiding errors, might contribute as well.

A clear picture about the fiber-injection losses involved is crucial for the conception of new observing programs, i.e. to estimate required exposure time and achievable S/N, but also for scheduling of planned observations, and during the design of new spectrographs to predict their expected performance and make the most appropriate design choices.
Also, one would like to correct raw spectra for fiber-injection losses to achieve a better spectrophotometric calibration.
All of this requires that an accurate description of the IQ and associated fiber-injection efficiencies is implemented in the exposure time calculator (ETC) and in the data reduction system.
The bottleneck here is that for spectrographs, and in particular fiber-fed ones, one typically has no simultaneous and accurate monitoring of the IQ during the observations. Without a proper estimate of this, any computation by the ETC and consequently any comparison between predicted and observed performance becomes highly uncertain and not particularly informative.

In the following, we conduct a thorough investigation of the IQ delivered to ESPRESSO, making extensive use of its acquisition and guiding cameras which provides ground-truth information about the received PSF.
We explore how representative the estimates derived from facility-provided seeing and IQ monitors are and how accurately these can predict the fiber-injection losses.
In addition, we present a novel method to infer the IQ profile from the integrated images of the ESPRESSO field-stabilization cameras.
These do provide, at least in principle, a simultaneous assessment of the IQ received by the instrument over the course of the spectroscopic observations. We quantify the precision and accuracy of this method and test whether these estimates provide a better characterization of IQ and fiber-injection efficiencies than the facility-provided values.

\section{Existing IQ Measurements at Paranal}

At the ESO Paranal observatory, numerous facilities exist to monitor the environmental conditions and provide contextual information to the user supplementing the actual science observations.

\subsection{DIMM-based IQ Model}
\label{Sec:IQ_DIMM}

The most relevant stand-alone device for measuring the atmospheric seeing is, arguably, the Differential Image Motion Monitor (DIMM, \citealt{Tokovinin2002, Sarazin1990, Chiozzi2016}).
During the night, it constantly measures the amount of atmospheric turbulence along the line-of-sight towards a reference star.
From the differential motion of two stellar images obtained through separate sub-apertures, and assuming Kolmogorov turbulence theory in which the spectral power density depends on the spatial length scale as $P(r) \propto r^{\frac{5}{3}}$, the characteristic coherence length-scale of the atmosphere, $r_0$, also called the \textit{Fried parameter} \citep{Fried1966}, is inferred. Within the same framework, this can be related to the full-width at half maximum (FWHM) of the blurring introduced to a long-exposed image of a point source as
\begin{equation}
\epsilon_0 = 0.976 \; \lambda / r_0,
\label{Eq:Seeing}
\end{equation}
where $\epsilon_0$ is commonly defined as the \textit{seeing} and $\lambda$ denotes the wavelength of observation.
Seeing measurements from the DIMM, normalized to a default reference wavelength of $500\,\textup{nm}$ and airmass zero, i.e. for observations at zenith, are available in real time and at high cadence via the ESO \textit{Astronomical Site Monitor} (ASM).
In addition, DIMM readings from the start and end of each observation are attached to the science data and listed in the \fits headers under the key words \texttt{ESO\:TEL\:AMBI\:FWHM\:START} and \texttt{ESO\:TEL\:AMBI\:FWHM\:END}.

Unfortunately, the seeing reported from the DIMM does not directly correspond to the IQ observed at the telescope. An important factor is that the  Kolmogorov model does in practice not accurately describe the atmosphere. More appropriate is the von~K\'arm\'an representation of the seeing, introduced by \citet{Tokovinin2002} and described in detail also by \cite{Martinez2010}. Here, an additional parameter, the \textit{outer scale}, $L_0$, describes up to which scale the turbulence spectrum follows the $\propto r^{\frac{5}{3}}$ law, but exhibits substantially reduced power beyond this. This leads to a modified description for the seeing-induced blurring of the form
\begin{equation}
\epsilon_\mathrm{vK} =  \epsilon_0 \; \sqrt{ \, 1 - 2.183 \; (r_0 / L_0 )^{0.356.} \, } .
\label{Eq:eps_vK}
\end{equation}
For the outer scale, \cite{DaliAli2010} report a median value at Paranal of $L_0=22\,\textup{m}$, while \citet{vandenAncker2016} instead advocate for a value of $L_0=46\,\textup{m}$ to better match the IQ observed with VISIR in the mid-IR. The ESO ETC as well adopts the latter value.
Apart from this, the FWHM of the seeing-induced blurring depends on the airmass ($AM$) and the wavelength of the observed target following
\begin{equation}
FWHM_\textup{Atm} = \epsilon_\mathrm{vK} \; \left( \frac{\lambda}{500\,\mathrm{nm}} \right)^{-0.2} AM^{0.6} .
\label{Eq:Airmass}
\end{equation}
Finally, the IQ might as well be affected by aberrations from the telescope which, in the simplest form, can just be added in quadrature, yielding the final DIMM-derived estimate for the IQ:
\begin{equation}
FWHM_\textup{DIMM} =\sqrt{ {FWHM_\textup{Atm}}^2 + {FWHM_\textup{Tel}}^2 } \; .
\label{Eq:FWHM_DIMM}
\end{equation}
This scheme is also adopted by the ESO Exposure Time Calculator (ETC)%
\footnote{ \url{https://www.eso.org/observing/etc/doc/helpuves.html} }, %
with the assumption of $FWHM_\textup{Tel}=0.2"$.

The constant availability of the DIMM seeing measurements is without doubt highly convenient, however, estimating the IQ from the DIMM measurements comes with several disadvantages. This includes the dependence on the various model assumptions, but also the fact that the seeing is inferred along a different sightline compared to the actual science observations.
The DIMM is located on its own tower in the northeastern part of the VLT platform \citep[e.g.][]{Griffiths2024}. Strongly localized effects in the ground-layer turbulence, e.g. due to height, topography, or nearby buildings, can affect the overall seeing measurement, which did manifests itself e.g. in the systematically larger seeing measured by the old DIMM (in operation between 1998 and 2016) that was placed at the northernmost edge of the VLT platform%
\footnote{ESO site-monitoring manual: \url{http://archive.eso.org/cms/eso-data/ambient-conditions/Astronomical_Site_Monitor_Data_User_Manual_v20171107.pdf}}.
Furthermore, the DIMM-derived IQ estimate can not capture possible dome-seeing or effects related to the tip-tilt correction of the VLT active field-stabilization loop.

\subsection{Guide Probe IQ Measurement}

Another, more direct type of IQ measurements is provided by the VLT Unit Telescopes (UTs) themselves. For regular operation of the UTs and in particular their active optics system, each focal station is equipped with a guide~probe, a movable arm that picks up the light of a selected guide~star. This, in principle, provides direct, simultaneous, real time information about the IQ as it is delivered by the telescope to the focal plane. Apart from a small offset of less than a few arcmin, the guide~star probes exactly the same path through the atmosphere as the science observation and is affected by the identical telescope-related aberrations as well as guiding and field-stabilization effects. The spectral energy distribution (SED) of the guide~star might deviate from that of the science target and the effective, bandpass-averaged IQ might therefore be different, but the wavelength-dependence of the atmospheric seeing is relatively weak, $\propto\lambda^{-0.2}$, and, since the SED of the guide star is known%
\footnote{but unfortunately not reported to the user by the ESO data flow},
in principle correctable. In addition, the measurements from the guide~probe are subject to differential atmospheric refraction%
\footnote{Except for the UT1 (and previously UT2) Cassegrain focal station, since the FORS\,2 (and previously FORS\,1) LADC prisms are actually located in the Cassegrain tower and therefore ahead of the guide probe: \url{http://www.eso.org/sci/facilities/paranal/instruments/fors/doc/VLT-MAN-ESO-13100-1543_P113.pdf}},
while the scientific instruments are usually equipped with an (independent) atmospheric dispersion corrector (ADC). This effect, however, could also be taken into account.
Since our interest here focuses on the IQ delivered to the scientific instrument and the associated fiber-injection losses, IQ measurements from the guide~star are conceptually better suited, compared to values derived from the DIMM or similar stand-alone facilities, that aim more on monitoring fundamental properties of the atmosphere above the observatory.
In practice, however, the UT guide probes and in particular the data flow is not really optimized to provide the user a high-fidelity monitoring of the IQ during the course of his observations.

In the design implemented at the VLT, the guide~probe splits the light of the guide~star with a dichroic beam-splitter%
\footnote{To our understanding, the dichroic cuts around $600\,\textup{nm}$ and sends the long-wavelength light to the AGS ($600$ to $700\,\textup{nm}$) and the short-wavelength one to the AOSH ($500$ to $600\,\textup{nm}$), see \citet{Martinez2012} and \url{https://www.eso.org/sci/facilities/develop/documents/VLT-SPE-ESO-10000-2723_is1.pdf}. In contradiction to this, the ESO site-monitoring manual states an effective wavelength for the AOSH of $700\,\textup{nm}$.}
and delivers it to two separate instruments, the actual guide camera (AGS) and a Shack-Hartmann wavefront sensor (AOSH).
The guide camera, used for field acquisition and guiding, takes images of the guide star and therefore directly sees the PSF delivered by the telescope to the instrument. It does this with relatively fine sampling and high S/N. However, the feedback loop for field stabilization, using the movable secondary mirror of the UTs (M2), runs fast (to our understanding up to $10\,\textup{Hz}$) and thus the exposure times for individual frames of the guide camera are short (\citealt{Martinez2012} state exposure times no longer than $50\,\textup{ms}$). Therefore, the PSF in individual guide~camera frames is not representative of the averaged IQ observed in long-exposed images, that we are interested in. Instead, the seeing is to a certain degree \textit{frozen}, an effect exploited e.g. by the \textit{lucky imaging} technique \citep[e.g.][]{Fried1978,Hippler2009}.
One could in principle take the individual guide~camera frames, stack them, compute averages over e.g. one minute, and measure the PSF with high fidelity on these stacked images. These, together with the derived quantities, could as time-resolved measurements be attached to the actual science data and made available to the user, representing the ideal and nearly perfect monitoring of the IQ. Unfortunately, such a concept has not been implemented in practice and in consequence, no IQ measurements from the guide~camera are made available%
\footnote{If we recall this correctly from our experience at Paranal, there are indeed some quality indicators derived from the guide camera images, but these are to our knowledge only available to the telescope operator in real time and cannot be accesses by the user after the observations have been completed.}.

Instead, ESO provides IQ measurements derived from the AOSH data. The primary function of the AOSH is to measure, at low cadence, residual wavefront errors at the telescope focal plane and thereby provide correction signals for the active optics system of the UTs, which utilizes about 150 actuators to bend the primary mirror into its ideal shape and adjust the alignment of the secondary mirror along multiple degrees of freedom, of which one corresponds to the telescope focus. The measurement of the wavefront error is based on the Shack-Hartmann principle \citep{Hartmann1904,Platt2001} and uses a micro-lenselet array creating nearly 450 sub-apertures.
To achieve sufficient S/N, and because no fast control loop is needed, the exposure time for the AOSH is relatively long.
\cite{Martinez2012} state an average exposure time of $45\,\textup{s}$, the ESO site-monitoring manual %
%
%
reports $20\,\textup{s}$. Anyway, both should be long enough to average-out the atmospheric turbulence and converge to the IQ seen in long-exposed observations. For the IQ measurement, the numerous images created by the individual sub-apertures are re-centered, stacked, and the FWHM of the averaged star image determined, following the procedure described by \citet{Martinez2012}.
Conceptually, the observed flux distribution is modeled by a function of form
\begin{equation}
F(r) =  \frac{ 10 \; {\log(2)}^{3/5} }{ 3 \; \Gamma(\frac{3}{5}) \;  \theta } \; \exp\left( -\log(2) \; \left( \frac{2 \; r}{\theta} \right)^{5/3} \; \right) \; ,
\label{Eq:FiveThirds}
\end{equation}
in which $r$ denotes the spatial coordinate, i.e. the radius for a rotationally-symmetric flux distribution. We parametrize this distribution here as function of the FWHM, $\theta$, include the proper normalization, and will in the following refer to it as the \textit{five-thirds distribution}.
However, being a Shack-Hartmann sensor designed to accurately measure the displacement of the individual spot images, the AOSH has a pixel sampling of only 0.305"/pix, which is less than ideal for quantifying the width and shape of the PSF. In addition, the size of the Shack-Hartmann sub-apertures corresponds to a diameter of only $33.8\,\textup{cm}$, which implies a diffraction limit of about $0.4"$. Both aspects severely complicate the IQ measurement, in particular under good seeing conditions, and \citet{Martinez2012} describe two different algorithms developed to \textit{de-convolve} the observed spot size measurement for the diffraction effect caused by the individual sub-apertures.
To our understanding, these two estimates for the FWHM of the true PSF are reported in the \fits headers under the key words \texttt{ESO\: TEL\:IA\:FWHMLINOBS} and \texttt{ESO\:TEL\:IA\:FWHMOBS}%
\footnote{We stress that the key word \texttt{ESO\:TEL\:IA\:FWHMOBS} is actually not present in the headers, although for completeness it should.}.
In addition, \texttt{ESO\:TEL\:IA\:FWHMLIN} and \texttt{ESO\:TEL\:IA\:FWHM} provide the same measurements scaled to unity airmass, i.e. corresponding to observations at zenith. The reported values (probably) represent the conditions at the start of the observation (i.e. right after acquisition). No time-series information are provided and thus changes in the seeing or corresponding IQ over the course of long exposures can not be monitored.

While conceptually measuring the IQ at the most appropriate location, the characteristics of the AOSH limit the accuracy of the derived IQ values, in particular due to the poor sampling and small apertures. Thus, despite the \textit{de-convolution} attempts described in \citet{Martinez2012}, the \texttt{ESO\:TEL\:IA\:FWHMLINOBS} key words on average overestimate the FWHM of the actual IQ. Supposedly, this offset depends on the focal station
used, but appropriate calibration coefficients, if ever determined, are not provided by ESO and therefore not available to the user.
Also, \citet{Martinez2012} stress that the AOSH IQ measurements would be independent of vibrations, guide errors, and the telescope field-stabilization (active tip-tilt correction). While this might be an advantage if one wants to characterize the atmosphere, these effects certainly influence the IQ delivered to the science instrument, which--at least conceptually--makes the AOSH-derived measurements unrepresentative for our purpose.

\subsection{ESPRESSO Acquisition and Field-Stabilization Camera}

In addition to these facility-provided seeing and IQ measurements, there exist additional possibilities within ESPRESSO to assess the IQ.
Since the spectroraph is located in the combined Coud\'e laboratory and to use it with any or all UTs simultaneously, the light is guided from the Nasmyth focus through the corresponding Coud\'e train to the basement of each UT and from there through underground tunnels to the incoherently combined Coud\'e focus (ICCF) where the spectrograph front-end injects the light into the spectrograph fibers \citep{Cabral2010, Cabral2012, Cabral2014, Megevand2014, Pepe2021}.
To correct for some (unavoidable) misalignments introduced along the rather long Coud\'e trains and to facilitate acquisition of the science target, ESPRESSO implements a secondary guiding scheme%
\footnote{Primary guiding by the UT is still performed at the Nasmyth focus.}.
The spectrograph front-end therefore incorporates an acquisition and field-stabilization camera that stares at the pierced mirror in which the spectrograph fibers are mounted and a corresponding small tip-tilt mirror \citep{Riva2014b, Pariani2016, GraciaTemich2018a}.
For acquisition, the target is identified and placed onto the spectrograph fiber using this camera. If desired by the user, an extra step is performed in which the target is deliberately offset from the fiber hole and a dedicated acquisition image is taken that allows to directly assess the IQ.
During the science integration, the field-stabilization camera observes the fiber-entrance hole and uses the spill-over light to adjust the guiding and keep the target centered on the fiber \citep{Riva2014b, Landoni2016, Calderone2016}.

\begin{figure*}
 \includegraphics[width=\linewidth]{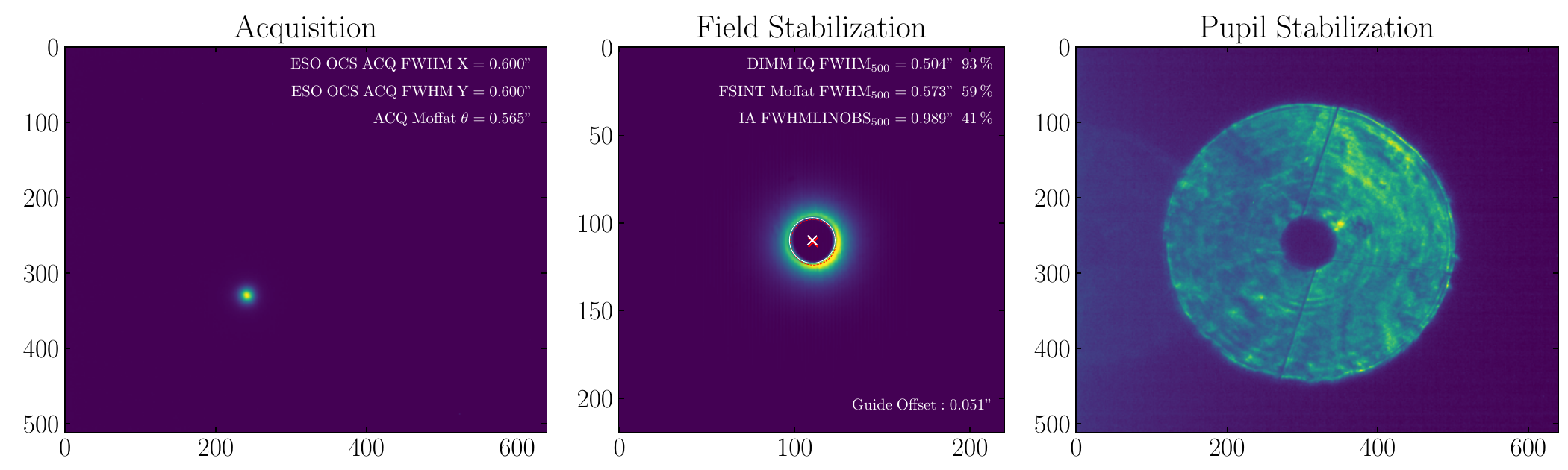}
 \caption{
  Example of acquisition, integrated field-stabilization, and pupil-stabilization image (left to right) attached to one selected ESPRESSO exposure (\texttt{ESPRE.2019-07-07T23:06:18.991}).
  The FWHM determined by the ESPRESSO data flow as well as our Moffat fit to the acquisition image are stated in the left panel.
  In the central panel, FWHM estimates and corresponding fiber-injection efficiencies, derived from the DIMM-based IQ model, our Moffat fit to the field-stabilization image, and the IQ measured by the AOSH sensor, re-scaled to a common wavelength of $500\,\textup{nm}$, are stated.
  The guide offset denotes the distance between our estimate for the location of the fiber hole (white cross and circle) and the location reported in the \fits headers (red cross and circle).
  The pupil-stabilization images are usually taken and attached to the spectroscopic data, but the control loop, acting on a corresponding tip-tilt mirror, is in fact not active.
 }
 \label{Fig:Guiding}
\end{figure*}

While nearly all spectrographs are equipped with some sort of slit-viewer and display those images in real-time at the instrument control station, it is relatively unique that for ESPRESSO these frames are stored and attached to the science data. This allows to assess IQ and guiding performance in retrospect, also when the observations were taken in service mode and the user was not present at the mountain. An example for the frames provided along the spectroscopic data is shown in Figure~\ref{Fig:Guiding}.
In case an acquisition frame was taken, the user receives an actual image of the science target attached to the spectroscopic data%
\footnote{This is provided in form of an extra \fits extension in the data file.}
and, given that the target is point-like, has a direct documentation of the PSF delivered to the ICCF focal plane at the time of acquisition%
\footnote{The ESPRESSO front-end contains a set of ADC prism to correct for the effect of differential atmospheric refraction.}.
Also, derived FWHM measurements of the target are included in the \fits headers%
\footnote{The corresponding key words are \texttt{ESO\:OCS\:ACQ1\:FWHMX\:ARCSEC} and \texttt{ESO\:OCS\:ACQ1\:FWHMY\:ARCSEC}, where the number indicated the UT used.}.
These are computed from two cuts through the peak along $X$ and $Y$ direction, not providing a third parameter needed to fully describe an elliptic two-dimensional profile, and assume a Gaussian shape, which is not able to describe seeing-broadened PSFs in a particularly accurate way. In addition, the exposure time of the ESPRESSO acquisition images is fixed to $4\,\textup{s}$, barely sufficient to average-out the atmospheric disturbances and be representative for long-exposed images. Nevertheless, the acquisition images are the best option to get an accurate assessment of the IQ delivered to the ICCF focal plane at the start of the observation.

In addition, the ESPRESSO data flow also provides information from its own field-stabilization loop to the user. Here, not every single frame taken over the course of the potentially long science integration is stored, but a stack, corresponding to the integrated light seen by the field-stabilization camera, is attached to the spectroscopic data%
\footnote{Again, in form of a separate \fits extension, labeled \texttt{FSINT}.}.
This allows to assess--at least qualitatively--whether the target was on average properly centered on the fiber hole (see Figure~\ref{Fig:Guiding}, central panel).
In principle, the integrated field-stabilization image also provides information about the average IQ during the observation.
The difficulty for deriving quantitative and accurate measurements from it is, of course, that in the given setup the core of the PSF profile is unobservable and only the residual spill-over light that is not sent into the spectrograph fiber is available for analysis. Determining the IQ shape only from its wings will certainly impose some limitations and be less precise than having access to the full profile, however, if possible to make work in a reliable fashion, IQ measurements derived from the stacked field-stabilization frames also come with several advantages:
first of all, since \texttt{FSINT} frames are delivered by default with every on-sky spectrum, they would be available for every ESPRESSO observation taken (past or future) and, being computed from existing data, would come for free without adding any overheads.
Also, the integrated field-stabilization images reflect the actual IQ received by the instrument, including atmospheric, telescope, and guiding effects, and are therefore most representative for determining fiber-injection losses. In addition, they capture, although not time-resolved but just as an average, the conditions over the full course of the science integration, even if these change.
Therefore, it is worth to investigate how well the IQ can be determined from the integrated field-stabilization images and whether this provides a more accurate measurement than the estimates derived from the DIMM or AOSH.

\section{Data Selection}
\label{Sec:Data}

Most important for a proper assessment of any IQ measurement strategy is a reference dataset that provides--as much as possible--\textit{ground truth} information to which one can compare. For our case, the most appropriate data in this regard are the ESPRESSO acquisition images.
Although they are taken with rather short exposure time, they provide the full PSF profile in exactly the way it is delivered to the instrument.
This allows, in contrast to numerical FWHM values inferred from other sources, to measure the full IQ profile instead of being forced to make assumptions about its shape, e.g. being Gaussian or following the \textit{five-thirds~distribution} (Equation~\ref{Eq:FiveThirds}).
For our analysis, we want to ensure that the IQ seen in the acquisition image and in the stacked field-stabilization frames represent the same atmospheric condition. Therefore, we require the exposure time of the science integration to be relatively short.
A natural choice are therefore the observations of spectrophotometric standard stars.
These calibrations are executed regularly but under various seeing conditions, target bright A-type stars ($m_\textup{V}<6\,\textup{mag}$) with corresponding integration times $\leq240\,\textup{s}$, and, being part of the standard instrument calibration plan, become public immediately.
In 2019, the standard template for all kinds of observations was changed to by default not take the acquisition image. This saves about 48 seconds during the acquisition process and increases the instrument efficiency slightly. Since then, the majority of observations have no acquisition image attached. However, following our request, observations of spectrophotometric standard stars are again taken with acquisition image since Summer 2022 and therefore provide a continuously growing resource for IQ studies.

From the ESO archive, we obtain in total \VAL{201} spectrophotometric standard star observations that were taken in \texttt{1HR1x1} mode%
\footnote{Calibrations for various instrument modes are typically taken in a joint observing block with only one acquisition procedure at the beginning. Thus, only the first spectrum, here the one in \texttt{1HR1x1} mode, contains the acquisition image.}
and contain an acquisition image. Unfortunately, not all exposures are usable for our purpose. In a significant fraction (\VAL{$\approx15\,\%$}) of acquisition images, the target star is saturated, rendering any reasonable IQ measurement futile. Furthermore, we notice that in numerous acquisition images the target appears disturbed, asymmetric, or with significant substructure. We suspect that this is a consequence of the rather short exposures, during which the atmospheric turbulence had insufficient time to properly average-out and yield a round and smooth flux distribution.
We manually inspect all acquisition images and reject those which appear clearly deformed.
In addition, there are a few cases for which the stacked field-stabilization image shows signs of saturation.
Excluding all those, we remain in total with \VAL{141} usable exposures. Numerous standard star observations were taken in Summer 2019, while basically no data with acquisition image exist for 2020 and 2021. During the other periods, spectrophotometric standard star observations are taken at moderate cadence, typically not more than once per month with each of the four UTs.
We stress that the ESPRESSO fiber feed has been exchanged in June 2019. Since then, the fiber hole seen by the field-stabilization camera has a notably different appearance, exhibiting substantially less scattered light from the inner wall of the hole.


\section{Data Fitting}

The cameras used in the ESPRESSO front-end  are of type Allied~Vision Bigeye G-132 Cool%
\footnote{\url{https://cdn.alliedvision.com/fileadmin/pdf/en/Bigeye_G-132_Cool_DataSheet_en.pdf}},
constructed around the Sony~ICX285 CCD sensor with 1280×1024 pixels and operated in 2×2 binning mode \citep[see][]{Landoni2016, Calderone2016}.
The plate scale is approximately $0.0416\,\textup{arcsec/pix}$.
We emphasize that these are commercial devices \citep{Duhoux2014}, which are quite sensitive and facilitate reliable guiding even on faint targets ($\lesssim19\,\textup{mag}$), but have not the same properties as a science-grade detector.
For instance, they operate with a dynamic range of just 12\,bits, i.e. a highest read value of $4095\,\textup{ADU}$.
Also, they are not characterized in the same way and calibrations typically taken with science detectors, like bias, dark, gain, flatfield, etc., are not available.
We therefore perform only a basic pre-processing of the frames, applied to the acquisition images and stacked field-stabilization frames in the same way. For this we, mask areas that contain flux from the target and compute the global median value from the empty area. In addition, we also determine the mean column and row profiles and subtract all three from the raw data. This effectively removes a rather pronounced pattern between even and odd columns. Since the gain value is not known, we can not compute proper Poissonian photon errors. For the field-stabilization frames, it is anyway a bit unclear which operations have been executed in the stacking procedure and how many frames have been averaged. Thus, no proper error vector can be established.

\subsection{Fitting of Acquisition Images}

The ESPRESSO data flow provides, along with the FWHM measurements, also the determined location of the target in the acquisition image.
This is typically very accurate, nevertheless, we re-fit the centroid of the profile, assuming a two-dimensional elliptical Gaussian profile. From this fit we only retain the coordinates of the center position for the subsequent steps and drop the other parameters, in particular the information about the shape.

After having determined the centroid, we bin the observed flux in circular annuli which provides the empirical radial profile of the PSF in one pixel wide bins.
The acquisition image likely contains sufficient information to infer a two-dimensional, non-rotational-symmetric profile. However, we want to apply the identical scheme also to the integrated field-stabilization frames. For these, given that the core of the profile is unobservable, it will already be a challenge to determine a meaningful width of the PSF assuming just a rotationally symmetric profile. Adding additional parameters would probably overstretch the information content. Thus, we restrict the analysis to symmetric PSF profiles.

We fit the empirically determined radial PSF profiles with an analytical function. For this, we use a Moffat profile of the form
\begin{equation}
F_{\alpha,\,\beta}(r)  = \frac{\beta-1}{\pi \; \alpha^2} \left( 1 + \frac{ r^2 }{\alpha^2} \right)^{-\beta} \; ,
\end{equation}
which, expressed as function of the FWHM, $\theta$, turns into
\begin{equation}
F_{\theta,\,\beta}(r)  =  \frac{4 \: \left( \beta - 1 \right) \: \left( 2^\frac{1}{\beta} - 1 \right) }{\pi \; \theta^2} \;
                        \left( 1 + \frac{ 4 \; \left(2^\frac{1}{\beta} - 1 \right) \; r^2 }{\theta^2} \right)^{-\beta}.
\label{Eq:Moffat}
\end{equation}
This functional form is, as already demonstrated by \citet{Moffat1969}, able to accurately describe the radial profile of a seeing-limited PSF.
The slight complication here, and maybe the reason why astronomers often resort to use inaccurate Gaussion or five-thirds distributions instead of proper Moffat profiles, is that the latter is described by two parameters and, in addition to the FWHM, also requires a shape parameter, $\beta$.
We perform the fit using a standard $\chi^2$-minimizer. Since no proper uncertainties for the flux in the images can be constructed, we assign empirical uncertainties to the binned PSF profile. For this, we compute the standard deviation of the flux in each annulus and divide it by the square-root of the number of contributing pixels.
In the limit of a rotationally-symmetric profile (and sufficient pixels in the annulus), this would provide the formally correct uncertainties.

\begin{figure}
 \includegraphics[width=\linewidth]{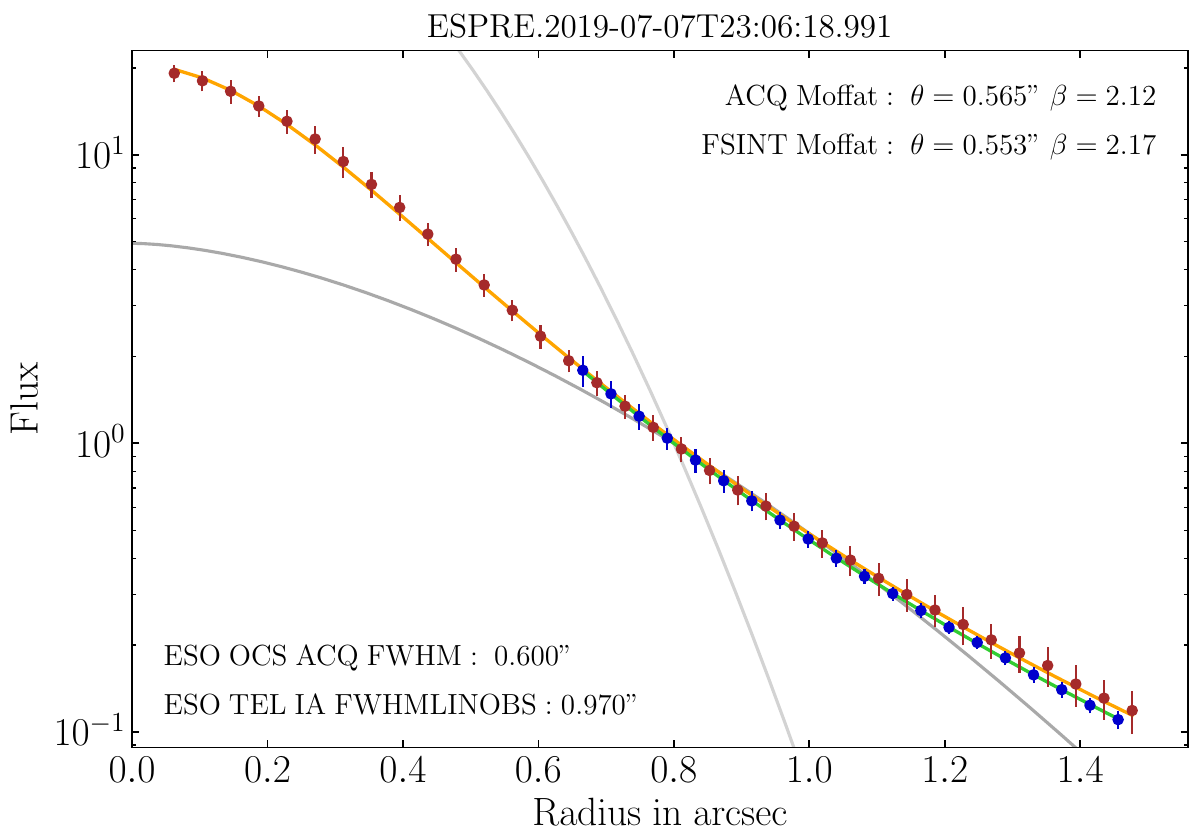}\vspace{4pt}
 \caption{
  Different radial flux profiles associated with one selected ESPRESSO exposure (the same as in Figure~\ref{Fig:Guiding}).
  Red datapoints represent the IQ profile extracted from the acquisition image (ACQ), while the orange curve indicates the Moffat distribution fitted to them.
  The profile from the integrated field-stabilization frames (FSINT) is shown by blue datapoints and the corresponding fit in green.
  A Gaussian profile (light gray) with the ESPRESSO acquisition image FWHM measurement and a five-thirds profile (dark gray) with the FWHM reported by the AOSH are shown as well.
  The relevant values from the file headers or obtained in the fits are stated in the figure.
  Flux units are fully arbitrary. Since the core of the field-stabilization profile is inaccessible,
  all datasets are re-scaled to exhibit the same flux at a radius of approx. 0.8".
  Uncertainties for the binned flux profiles are approximated in an empirical way and exaggerated by a factor of 10$\times$ for visualization.
 }
 \label{Fig:Profile}
\end{figure}

The binned radial flux profile for one selected acquisition image
is shown in Figure~\ref{Fig:Profile} together with the corresponding Moffat profile fitted to the data. Obviously, the analytic function is able to describe the observed empirical profile with a very high degree of accuracy.
In addition, Figure~\ref{Fig:Profile} shows a Gaussian profile that corresponds to the FWHM derived by the ESPRESSO data flow from the acquisition image%
\footnote{Here, we take the geometric mean of the values reported by the \texttt{ESO\:OCS\:ACQ1\:FWHMX\:ARCSEC} and \texttt{ESO\:OCS\:ACQ1\:FWHMY\:ARCSEC} key words to obtain an equivalent rotationally-symmetric profile.}
and a five-thirds profile (Equation~\ref{Eq:FiveThirds}) that represents the AOSH-derived IQ%
\footnote{Based on the \texttt{ESO\:TEL\:IA\:FWHMLINOBS} key word.}.
Clearly, these profiles do not properly describe the observed flux distribution. Here, the mismatch is not a question of the particular FWHM assumed. In principle, one could re-fit the functions to the empirical data but would still be unable to properly describe the observed profile, simply because the flux does not follow the assumed analytic forms.
Thus, any attempt to describe the observed IQ, or related quantities like the fiber-injection efficiency, in a truly accurate way while assuming a Gaussian or five-thirds profile is futile right from the start. This has already been shown by \citet{Moffat1969}, mentioned as well by e.g. \citet{Martinez2010}, and demonstrated again here.

\begin{figure}
 \includegraphics[width=\linewidth]{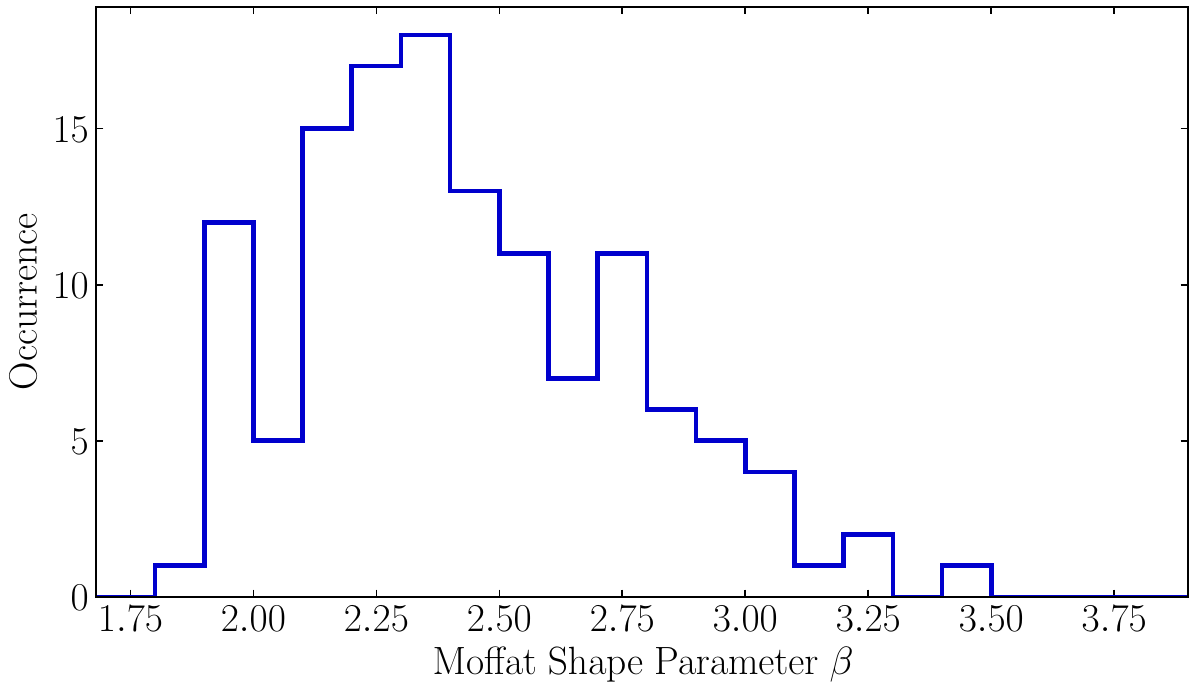}
 \caption{
 Histogram of the shape parameter, $\beta$, obtained from the Moffat fits to the acquisition images.
 }
 \label{Fig:Hist_beta}
\end{figure}

Figure~\ref{Fig:Hist_beta} shows a histogram of the determined $\beta$ parameters for the full set of acquisition images, but excluding exposures with determined FWHM of $\theta>1.5"$. The distribution peaks at $\beta\approx2.3$ and spans values from $1.8$ to $3.6$.

\subsection{Fitting of Field-Stabilization Images}

For the integrated field-stabilization frames, we follow in principle the same procedure as described above for the acquisition image, but of course restricted to regions outside the fiber hole.
Rather important here is to define the center of the intrinsic PSF profile, which is by itself not observable but disappears in the spectrograph fiber.
The ESPRESSO instrument control system regularly determines the location of the fiber hole and uses this as target position for the field-stabilization loop \citep{Landoni2016}. The last set of assumed fiber coordinates are reported in the \texttt{ESO\:OCS\:AG\:FIBERX} and \texttt{ESO\:OCS\:AG\:FIBERY} header key words.
We made our own attempts to fit the fiber hole in the field-stabilization images and, arguably, managed to determine it with better precision%
\footnote{The offset between the hole center position reported by ESPRESSO and our determination is on average $0.05"$, about one pixel.}.
However, for fitting the PSF profile, we achieved the best results by adopting the coordinates reported in the \fits headers.
This can be understood by considering that the center of the flux distribution is defined by the location towards which the field-stabilization loop steers the target, and not necessarily given by the actual position of the hole.

From the assumed center of the profile, we mask everything within $15\;\textup{pixel}$ radius%
\footnote{For data taken before June 2019, we adopt a radius of $16\:\textup{pixels}$.}
and compute the average flux in one pixel wide annuli, with the first bin center at a radius of $15.5\:\textup{pixels}$, corresponding to approximately $0.645"$, and the outermost one at $36.5\,\textup{pixels}$ ($1.52"$).
Here, we stress that the overall performance of the fit depends quite substantially on the choice of these values, in particular the inner limit. Utilizing data from smaller radii is in principle desirable to sample more of the core of the profile.
At the same time, the hole itself has to be safely excluded, as well as flux coming from the distorted edge of the drilled hole which does not follow the distribution of the actual PSF profile \citep[see][for a description of the drilling and polishing process]{GraciaTemich2018a}. We therefore performed quite extensive optimization attempts and finally settled on the described values.
After extraction of the empirical flux profile, we again fit the data with a Moffat profile as described above. To stabilize the fit, we impose a mild prior on the shape parameter, $\beta$.  Following the statistics obtained from the acquisition images (Figure~\ref{Fig:Hist_beta}), we adopt a Gaussian prior with a mean of 2.3 and a standard-deviation of 0.2. This helps to avoid large excursions but does not bias the result by a significant amount.

The flux profile obtained from the field-stabilization image and the corresponding Moffat fit are also shown in Figure~\ref{Fig:Profile}.
In this particular example, the profile extracted from the field-stabilization image follows rather closely the outer part of the profile derived from the acquisition image. In consequence, the fitted model parameters (FWHM and $\beta$) are nearly identical, demonstrating that an IQ measurement from the integrated field-stabilization image is indeed possible. The fitted FWHMs are also indicated in Figure~\ref{Fig:Guiding}.

\section{ Performance Evaluation}

While Figure~\ref{Fig:Profile} demonstrates that at least in individual cases one can obtain IQ measurements from the integrated field-stabilization images, a statistical analysis is needed to assess if this is possible in general and which precision and accuracy one can expect.
For this, we make use of the extensive sample of \VAL{141} usable exposures collected for this study. However, we exclude \VAL{12} observations for which the determined FWHM in the acquisition image is $>1.5"$.
At seeing conditions substantially worse than this, spectroscopic observations with ESPRESSO are anyway not particularly tempting, given the enormous fiber-injection losses ($\gtrsim80\,\%$).

\subsection{Comparison of Inferred FWHM}

In the following, we consider the Moffat-fitted FWHM measurements from the acquisition images as the reference that is closest to the truth and to which we compare other measurements.
In principle, one could also derive the FWHM in a non-parametric fashion directly from the observed profile. However, as demonstrated in Figure~\ref{Fig:Profile}, the assumed Moffat function is able to describe the observed profile with great accuracy. Thus, the FWHM determined in the fit should be virtually identical to any empirical FWHM measurement and we therefore stick with the parametric estimate.
To ensure a fair comparison, we convert all IQ measurements to a reference wavelength of $500\,\textup{nm}$, following Equation~\ref{Eq:Airmass}. The effective  wavelengths of the observations, however, have to be estimated. For ESPRESSO, a dichroic sends the near-IR light (probably $\lambda > 800\,\textup{nm}$) to the pupil-stabilizing camera and the visible light to the field-stabilization camera \citep{Landoni2016}.
Based on the efficiency curve provided by camera manufacturer, the dichroic, and the spectral energy distribution of the observed spectrophotometric standard stars (all A-type), but ignoring other contributions, we estimate the effective wavelength to be around $520\,\textup{nm}$.
In the guide probe of the UTs, light with wavelengths between $500$ and $600\,\textup{nm}$ is fed to the AOSH \citep{Martinez2012} and we thus assume an effective wavelength of $550\,\textup{nm}$.
Here, some uncertainty arises from the unknown spectral type of the guide star, but the dependence on the effective wavelength is weak and does not pose a relevant source of uncertainty.
Seeing measurements from the DIMM are already reported for a reference wavelength of $500\,\textup{nm}$. Here, however, we apply the model defined by Equations~\ref{Eq:eps_vK} to \ref{Eq:FWHM_DIMM} to estimate the IQ at the actual airmass.

\begin{figure}
 \includegraphics[width=\linewidth]{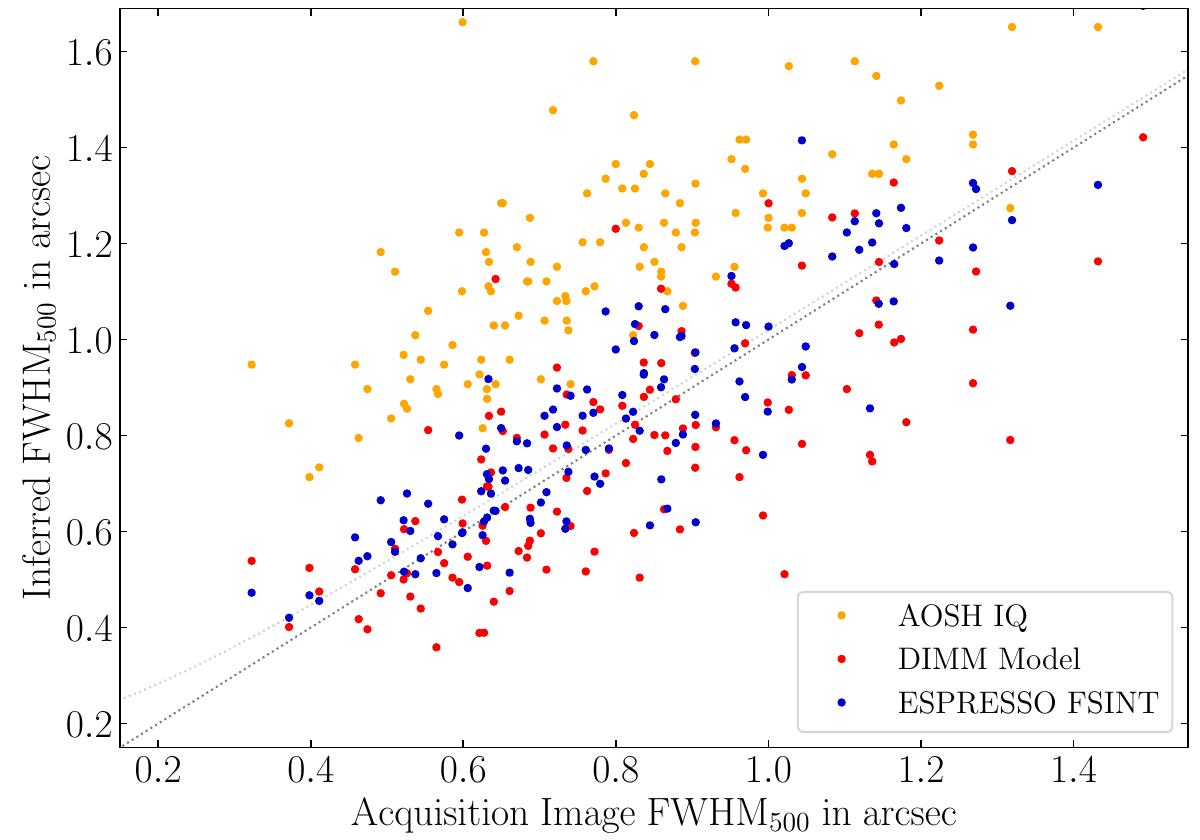}\vspace{4pt}
 \caption{
  Comparison of different IQ measurements to the FWHM observed in the ESPRESSO acquisition images.
  Colors indicate FWHMs inferred from the AOSH sensor in the guide probe of the UTs (yellow), a model for the IQ driven by the DIMM seeing measurements (red), and the ESPRESSO integrated field-stabilization frames (FSINT, blue).
  All measurements are re-scaled to a reference wavelength of $500\,\textup{nm}$. The dotted gray line indicates the ideal 1:1 relation. The light gray line assumes an additional guiding error of $0.2"$ that was added in quadrature.
 }
 \label{Fig:StatisticFWHM}
\end{figure}

Figure~\ref{Fig:StatisticFWHM} shows the comparisons of three different IQ estimates against the FWHM we derive from the acquisition images.
Obviously, the IQ derived from the AOSH systematically overestimates the FWHM by a quite substantial amount, i.e. \VAL{$+0.45"$} on average.
The DIMM-derived IQ very slightly underestimates the FWHM (by about \VAL{$-0.02"$}) and exhibits some noticeable scatter. In addition, there are four cases in our sample for which no valid DIMM value is reported.
The most accurate (in terms of bias) and most precise (in terms of scatter) estimate for the IQ during the science observation comes indeed from the analysis of the integrated field-stabilization images. We find a bias of just \VAL{$+0.03"$} and a standard deviation of \VAL{$0.11"$}.
This compares favorably to the precision of the other two methods. For the DIMM-derived FWHM, we find a (bias-removed) standard-deviation of \VAL{$0.17"$} and \VAL{$0.29"$} for the IQ from the AOSH.

Nevertheless, an IQ estimate with a RMS scatter of \VAL{$0.11"$} is in principle less precise than one would desire.
We therefore explored correlations with other factors to identify potential bottlenecks. However, no striking correlation, e.g. with the $\beta$~parameter, the airmass, exposure time of the spectrum,  or the date of observation was found. Also, there is no dependence on the UT used for the observation%
\footnote{The ESPRESSO front-end contains separate acquisition and field-stabilization cameras for each Coud\'e train coming from the four different UTs, but the pierced mirror and fiber assembly is always the same.}.

However, we did realize that the appearance of the field-stabilization image does depend on the details of the guiding procedure. For instance, in the example shown in Figure~\ref{Fig:Guiding}, the location of the fiber hole assumed by the ESPRESSO instrument control system (marked with a red cross) is slightly offset to the bottom-right compared to our estimate of the hole center (white cross). One also notices that the bottom-right rim of the hole appears brighter than e.g. the top-left one.
Visually inspecting many field-stabilization images, one gets the impression that the offset of the adopted hole centers indeed correlates with the brightness of the rim. This might suggest that the field-stabilization loop not always steers the target towards the optimal fiber center position. Despite this, we find no evidence indicating that our FWHM estimate would correlate with the magnitude of this offset, at least not as long as the center coordinates reported by ESPRESSO are used for the modeling.

From our personal experience at the observatory, we noticed that the field-stabilization loops (the combined effect of primary and secondary guiding) do not keep the target perfectly stationary on the fiber tip. Instead, the target wiggles around the desired position by a certain amount, which can also be seen in Figure~5 of \citet{Pepe2021}. The IQ in the integrated field-stabilization image therefore reflects the combined effect of atmospheric blurring and guiding errors. The latter effect might be much less pronounced in the acquisition images, due to the short exposure time of only $4\;\textup{s}$.
If we add in quadrature to the measured acquisition FWHM an assumed guiding error of $0.2"$ (indicated by a dashed gray line in Figure~\ref{Fig:StatisticFWHM}), the bias in the FWHM derived from the field-stabilization images vanishes completely. Although a plausible effect, it is fairly small and does not change the overall picture in any significant way.

In absence of any obvious cause for the scatter seen in Figure~\ref{Fig:StatisticFWHM}, we have to attribute this to stochastic effects. We stress, however, that the comparison between FWHM estimates obtained from the acquisition image and the integrated field-stabilization image contains noise from both measurements. The acquisition image should be affected only negligibly by photon noise, but the relatively short exposure time of just four seconds could make the IQ measurement itself noisy. Unfortunately, we can not quantify how much scatter visible in Figure~\ref{Fig:StatisticFWHM} is contributed by the acquisition image, but it could be possible that the IQ assessment from the field-stabilization images is actually more precise than total scatter seen in Figure~\ref{Fig:StatisticFWHM}.

\subsection{Fiber Injection Efficiency}

\begin{figure}
 \includegraphics[width=\linewidth]{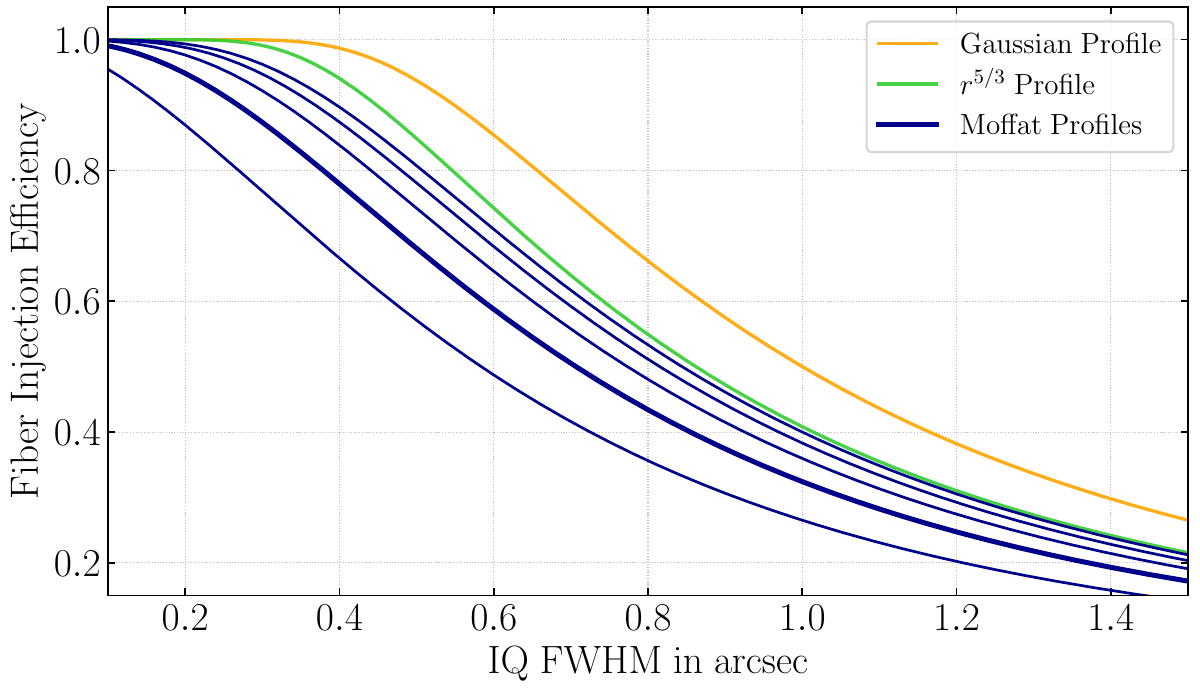}\vspace{4pt}
 \caption{
  Geometrical light-injection efficiency into a $1"$-diameter fiber as function of the IQ. A Moffat profile is assumed for the computation of the blue curves, which correspond to various profile shapes, bracketing the observed range (Figure~\ref{Fig:Hist_beta}) from $\beta=1.8$ (leftmost curve) to $\beta=3.8$ (rightmost) in steps of size 0.5.
  Orange and green curves visualize the injection efficiency assuming a Gaussian or five-thirds profile, respectively.
 }
 \label{Fig:InjectionEfficiency}
\end{figure}

One of our main interests within this study is to assess the fiber-injection efficiency. Having determined the IQ profile, this is readily possible.  When only considering the geometric efficiency and assuming rotational symmetry%
\footnote{The acceptance of incoming light by an optical fiber might in reality be more complex and e.g. depend on the angle of incidence. There are also losses due to reflections at the air-to-glass surface and ESPRESSO fibers are in fact octagonal.},
one simply has to integrate the two-dimensional IQ profile up to the radius of the fiber.
Since the Moffat function is an adequate representation of the actual profile, this can be done analytically.
For a given fiber diameter $D_\textup{f}$ and shape parameter $\beta$, the transmission as function of the IQ FWHM, $\theta$, can be expressed as
\begin{equation}
T_\mathrm{D_\textup{f},\,\beta}(\theta) = 1 - \left( 1 + \frac{ \left(2^\frac{1}{\beta} - 1 \right) \; {D_\textup{f}}^2 }{\theta^2} \right)^{1-\beta}.
\label{Eq:MoffatIntegral}
\end{equation}
This is visualized in Figure~\ref{Fig:InjectionEfficiency}. For several values of $\beta$, covering the range observed in the acquisition images (see Figure~\ref{Fig:Hist_beta}), the injection efficiency into a circular fiber with 1"~diameter is given as a function of the IQ.
One can easily see that the FWHM of the PSF needs to be smaller than 0.68" to ensure an injection efficiency better than $50\,\%$, if one assumes an average $\beta=2.3$. Over the full range of shapes, i.e. from $\beta=1.8$ to $\beta=3.8$, this might vary between $0.57"$ and $0.8"$.
Capturing more than 75\,\% of the light in the fiber requires an extremely good IQ between 0.38" an 0.57".
This highlight the massive impact of the fiber-injection losses on the overall throughput of instruments like ESPRESSO and the strong dependence on the atmospheric conditions.

Similarly to the description above, the fiber injection efficiency can also be computed for a Gaussian profile. In that case, the throughput is given by
\begin{equation}
T_\mathrm{D_\textup{f}}(\theta) = 1 - \exp\left( \; -\frac{1}{2} \; \left( \frac{ \sqrt{2\log(2)} \; D_\textup{f} }{\theta } \right)^2 \; \right) \; ,
\label{Eq:GaussIntegral}
\end{equation}
and for a five-thirds profile as
\begin{equation}
T_\mathrm{D_\textup{f}}(\theta) = \frac{ \gamma\left( \frac{6}{5}, \log\left(2\right) \: \left( \frac{D_\textup{f}}{\theta} \right)^\frac{5}{3} \right)}{ \Gamma\left( \frac{6}{5} \right) } \; ,
\label{Eq:FiveThirdsIntegral}
\end{equation}
where $\gamma(a,z)$ denotes the lower incomplete gamma function and $\Gamma(a)$ the gamma function, which are often combined into the regularized lower incomplete gamma function, $P(a,x) = \frac{\gamma(a,x)}{\Gamma(a)}$, further simplifying the notation.

Efficiency curves for a Gaussian and five-thirds function are shown as well in Figure~\ref{Fig:InjectionEfficiency}, highlighting the significant differences between the profiles.
The Moffat profile has more extended wings than a Gaussian. Thus, for profiles with the same FWHM, the injection efficiency is substantially lower for a Moffat profile. The difference is less pronounced with respect to the five-thirds profile, but still present, in particular for $\beta \ll 3.5$.
For a FWHM that equals the fiber diameter (in Figure~\ref{Fig:InjectionEfficiency} set to $D_\textup{f}=1"$), the Gaussian profile has a transmission of $50\,\%$. Under the same conditions, the five-thirds profile predicts an injection efficiency of just $41\,\%$, while this drops for the Moffat profile to $32\,\%$, assuming $\beta=2.3$, and might even be as low as $27\,\%$ for the extreme case of $\beta=1.8$.
Therefore, the actual shape of the IQ profile has to be carefully taken into account when predicting fiber-injection efficiencies and a substantial underestimation of the losses might result when using simplistic Gaussian distributions.

\begin{figure}
 \includegraphics[width=\linewidth]{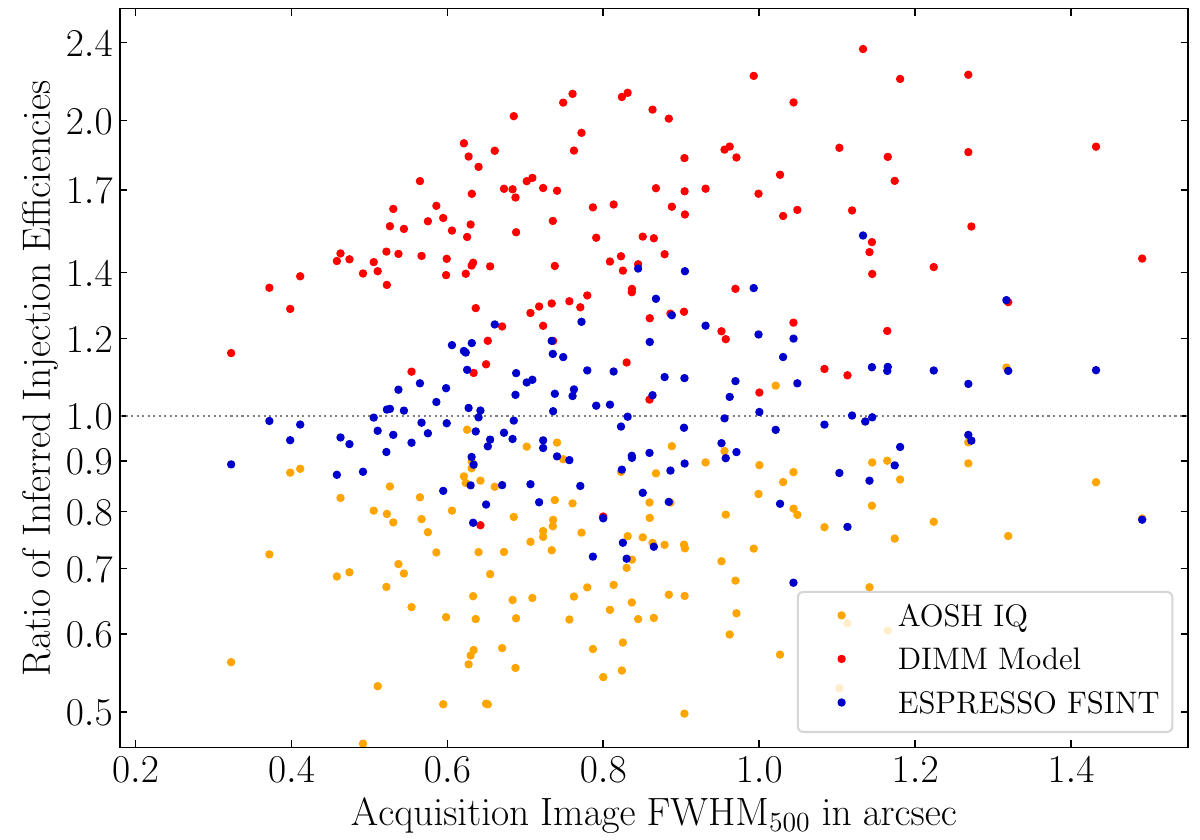}\vspace{4pt}
 \caption{
  Ratios between the fiber-injection efficiencies inferred from different IQ measurements and the one corresponding to the PSF profile observed in the ESPRESSO acquisition images.
  Colors indicate efficiency ratios corresponding to IQ from the AOSH sensor (yellow), a model driven by the DIMM seeing values (red), and the ESPRESSO integrated field-stabilization frames (FSINT, blue).
  All measurements are re-scaled to a reference wavelength of $500\,\textup{nm}$ and plotted against the FWHM measured in the acquisition images. The dotted line indicates the ideal 1:1 relation.
 }
 \label{Fig:StatisticThroughput}
\end{figure}

In the following, we compare the fiber injection efficiencies derived from the different IQ measurements.
As reference, we take again the Moffat fits applied to the acquisition images and compute the injection efficiencies that would be associated with the observed profiles, using Equation~\ref{Eq:MoffatIntegral}. The same formula is applied to the FWHMs derived from the field-stabilization images.
For the DIMM-based IQ estimate we compute, in accordance with the general concept lined out in Section~\ref{Sec:IQ_DIMM}, the fiber-injection efficiency based on a Gaussian profile (Equation~\ref{Eq:GaussIntegral}) while the IQ measurements from the AOSH are based on the approach described by  \citet{Martinez2012} and therefore adopt the five-thirds profile (Equation~\ref{Eq:FiveThirdsIntegral}).
To asses how well these metrics describe the fiber-injection efficiency, we compute the ratios relative to the value from the acquisition image and show the result in Figure~\ref{Fig:StatisticThroughput}.

One easily notices that the DIMM-based IQ model substantially overestimates the fiber-injection efficiency (by about \VAL{$52\,\%$}), despite providing rather accurate FWHM estimates (Figure~\ref{Fig:StatisticFWHM}). The reason for this is the inaccurate assumption of a Gaussian PSF that leads to a massive underestimation of the fiber-injection losses.
For the AOSH-derived IQ measurements, the overestimation of the fiber-injection efficiency by the five-thirds profile is much less severe, but here the FWHMs are already highly overestimated (Figure~\ref{Fig:StatisticFWHM}), which in consequence leads to an underestimation of the fiber-injection efficiency by about \VAL{$30\,\%$}.
The fiber-injection efficiencies derived from the integrated field-stabilization images, however, reproduces on average the ones derived from the acquisition images with great accuracy (formal bias \VAL{$<1\,\%$}) and exhibit a scatter of \VAL{$15\,\%$}.
The scatter for the other two types of estimates is, after removing the corresponding mean bias, similarly about \VAL{$23\,\%$}.
This demonstrates that accurate and relatively precise estimates of the fiber-injection efficiency can be derived from the integrated field-stabilization images.

One might consider to use as well a Moffat distribution to calculate the AOSH and DIMM fiber-injection efficiencies.
At least for the latter one, this should lead to less-biased estimates. However, the processes that deliver these measurements inherently assume a given profile shape (Gaussian or five-thirds) and simply using the provided FWHM value and plugging it into a totally different function does not seem particularly consistent. Also, the Moffat function requires a shape parameter, for which one could at best use some average value.
Apart from this, we stress that the estimate from the field-stabilization images does exhibit a smaller relative scatter than the other estimates. We therefore consider the analysis of the integrated field-stabilization image as the best and most appropriate method to estimate the IQ profile during the observations and to correct ESPRESSO spectra for fiber-injection losses.

\section{Conclusions}

In this study we have investigated the properties of the seeing-limited PSF delivered to the ESPRESSO spectrograph and explored several different methods to measure or predict the IQ. The goal here was not a characterization of atmospheric turbulence properties but instead an accurate assessment of the actual PSF width and shape, with a particular focus on the associated fiber-injection losses and their contribution to the overall instrument efficiency.
A relatively unique feature of the ESPRESSO data flow is that the acquisition and guiding images are stored, attached to the spectroscopic data, and made available to the user though the ESO archive (see Figure~\ref{Fig:Guiding}), providing valuable extra information.
From the acquisition images, an accurate assessment of the PSF profile delivered to the ICCF focal plane can be derived.
Predictions for the IQ based on various other sources, i.e. the DIMM or the AOSH sensor in the guide probe, have been compared and benchmarked against this reference.
In addition, we developed a scheme to infer the average IQ profile during the spectroscopic observation from the integrated field-stabilization images. Here, the spillover-light that is not injected into the fiber is used to inform about the PSF shape.

We demonstrated that reliable IQ measurement can be inferred using this method and that these estimates are more precise and more accurate than the predictions derived from the DIMM seeing or the AOSH sensor, in terms of FWHM as well as fiber-injection efficiency (Figure~\ref{Fig:StatisticFWHM} and \ref{Fig:StatisticThroughput}).
The measurements from the integrated field-stabilization images exhibit a negligible bias (\VAL{$+0.03"$} for the FWHM and \VAL{$<1\,\%$} for the injection fiber-efficiency) and a scatter of \VAL{$0.11"$} and \VAL{$15\,\%$}, respectively.
In comparison, the AOSH overestimates the FWHM on average by about \VAL{$0.45"$}, while the IQ model derived from the DIMM seeing is quite accurate (bias of \VAL{$-0.02"$}) but exhibit a larger scatter of \VAL{$0.17"$}.
In terms of fiber-injection efficiency, the DIMM-derived model substantially overestimates the efficiency by \VAL{$52\,\%$}, while the AOSH underestimates it by \VAL{$30\,\%$}, and both exhibit a (bias-removed) scatter of about \VAL{$23\,\%$}.
Here, the choice of a suitable function for modeling the IQ shape is crucial.
We adopt a Moffat profile and demonstrate that it describes the data with great accuracy.
The Gaussian profile (underlying the DIMM model) and five-thirds distribution (used for the AOSH measurement), however, are unable to accurately describe the PSF profile (Figure~\ref{Fig:Profile}) and therefore inevitably lead to incorrect predictions for the fiber-injection efficiency, even when the FWHM is correct.
We also stress that the fiber-injection losses are in general substantially larger than predicted by a Gaussian profile (Figure~\ref{Fig:InjectionEfficiency}). This can lead to overoptimistic estimates for the performance of an instrument or the S/N in  observations. When assuming a Moffat profile with a representative shape, even FWHMs substantially smaller than the fiber diameter, e.g. a FWHM of $0.7"$ for a $1"$~fiber, imply that more than $50\,\%$ of the light can not be accepted by the fiber and is simply lost.
This highlights the huge impact of fiber-injection losses on the overall efficiency of the spectrograph, the strong dependence on the atmospheric conditions, and underlines that the shape of the IQ profile is as important as its FWHM.

The general scheme presented here could in principle be implemented for many other instruments, since most fiber-fed spectrographs are equipped with a broadly similar fiber-viewer system.
However, very few of them actually provide frames from the field-stabilization camera to the user alongside the science data. The only other example that we are aware of is
NIRPS, the Near-InfraRed Planet Searcher \citep{Wildi2022}.
NIRPS, however, is fed by an adaptive-optics system which makes the results presented here not directly applicable and warrants a dedicated study.

Utilizing only the spill-over light that did not get injected into the fiber clearly has limitations. Since the most informative part of the IQ profile is unobservable, one has to resort to a relatively simple model that cannot capture all relevant effects, e.g. asymmetry of the PSF.
Also, the scatter of \VAL{$0.11"$}, although more precise than other estimates, is not entirely satisfactory.
In the future, spectrographs with a slightly different front-end design should allow for a much more accurate monitoring of the IQ.
For instance, in the current design of ANDES \citep[ArmazoNes high Dispersion Echelle Spectrograph,][]{Marconi2022}, the planned high-resolution spectrograph for the ELT \citep{Padovani2023}, the field-stabilization cameras do not stare at a pierced mirror. Instead, the light passes through a beam-splitter that sends a small fraction (probably $<1\%$) of the light towards the guide-camera while the rest goes to the fiber entrance. In this way, the guide-camera receives an image of the full PSF that should allow to characterize the IQ profile and the associated fiber-injection losses with great accuracy, probably even in a non-parametric fashion without making any assumption about the underlying profile. This would represent a natural evolution of the scheme we presented here.

For the moment, we believe that we have a method at hand that delivers improved IQ measurements for ESPRESSO and preliminary tests suggest that using our estimate of the fiber-injection efficiency derived from the integrated field-stabilization images to correct the raw flux in the spectra does indeed provide satisfactory results within the expected precision.

\section*{Acknowledgements}

We would like to thank Pedro Figueira for providing valuable insights into the various seeing and IQ measurements available at Paranal.
We also thank the ESPRESSO instrument operation team for re-instating the FWHM measurements for spectrophotometric standard star observations.
This research has made use of Astropy, a community-developed core Python package for Astronomy \citep{Astropy2013,Astropy2018}, and Matplotlib \citep{Hunter2007}.
TMS  acknowledgment the support from the SNF synergia grant CRSII5-193689 (BLUVES).
This work has been carried out within the framework of the National Centre of Competence in Research PlanetS supported by the Swiss National Science Foundation.

\section*{Data Availability}

All data used for this study is publicly available from the ESO archive facility: \url{http://archive.eso.org/cms/data-portal.html}.


\bibliographystyle{mnras}
\bibliography{./Literature}


\label{lastpage}

\end{document}